\documentclass[aps,prl,twocolumn,amsmath,amssymb,superscriptaddress,hidelinks]{revtex4-1}

\usepackage{bm}
\usepackage{verbatim} 

\usepackage{graphicx}
\usepackage[usenames,dvipsnames]{color}
\usepackage[colorlinks]{hyperref}
\hypersetup{
  colorlinks,
  citecolor=blue,
  linkcolor=blue,
  urlcolor=blue}

\renewcommand{\vec}[1]{{\bf #1}}

\usepackage[normalem]{ulem}

\begin{document}
\title{Quantum geometry induced microwave enhancement of flat band superconductivity}
\author{Arpit Arora}
\email{arpit22@ucla.edu}
\affiliation{Division of Physical Sciences, College of Letters and Science, University of California, Los Angeles (UCLA), Los Angeles, CA, USA}
\affiliation{Department of Electrical and Computer Engineering, UCLA, Los Angeles, CA, USA}
\author{Jonathan B. Curtis}
\affiliation{Division of Physical Sciences, College of Letters and Science, University of California, Los Angeles (UCLA), Los Angeles, CA, USA}
\affiliation{Institute for Theoretical Physics, ETH Z{\"urich, Z\"urich, CH}}
\author{Prineha Narang}
\email{prineha@ucla.edu}
\affiliation{Division of Physical Sciences, College of Letters and Science, University of California, Los Angeles (UCLA), Los Angeles, CA, USA}
\affiliation{Department of Electrical and Computer Engineering, UCLA, Los Angeles, CA, USA}

\begin{abstract}
Photo-control of correlated phases is central to advancing and manipulating novel functional properties of quantum materials. Here, we explore microwave enhancement of superconductivity in flat bands through generation of nonequilibrium quasiparticles at subgap frequencies. In conventional superconductors, it is known to occur via radiation absorption determined by fermi velocity, which however is small in flat bands resulting in quenched quasiparticle excitations. Strikingly, in contrast to the conventional paradigm we show a non-vanishing microwave absorption in flat band systems enabled by Bloch quantum geometry leading to superconducting gap enhancement, underscoring the band-geometric origin of nonequilibrium flat band superconductivity. Specifically, we demonstrate this in twisted bilayer graphene, a promising candidate material, and find significant gap enhancement near critical temperature. This work highlights that the nonequilibrium dynamics of materials with non-trivial flat bands as a promising area for future experimental and theoretical investigation. 
\end{abstract}

\maketitle

Flat band systems have led the way in exploration of unconventional phenomena in quantum materials with discoveries of superconductivity, spontaneously generated symmetry broken states and heavy fermion physics, most notably in graphene heterostructures~\cite{balents2020superconductivity,pablo2018tbgsc,yankowitz2019tuning,kim2022evidence,zhou2021superconductivity,sharpe2019emergent,serlin2020intrinsic}. 
The unique properties of these systems are driven by 
strong correlations among slow moving electrons and non-trivial winding of Bloch wavefunctions~\cite{balents2020superconductivity,flatband-sc-review}. In superconductors, this has led to a quantum geometric origin of supercurrent with topologically bounded superfluid weight~\cite{verma2021optical,torma-sc,bernevig-topologicalbound-sc} and coherence length~\cite{lau2023tbgsc,ktlaw2024ginzburg}, simulating great interest in flat band superconductivity. 
However, despite the great interest in these systems, their nonequilibrium properties have only just begun to be explored~\cite{Pyykkonen.2023,Shavit.2024}.

Here, we consider the effect of microwave radiation on flat band superconductivity, and in particular show it is possible to use microwave radiaiton to realize a dramatic nonequilibrium enhancement of superconductivity in flat band materials.
In conventional superconductors, microwave-enhanced superconductivity was first discovered by Wyatt~\cite{Wyatt.1966} and Dayem~\cite{Dayem.1967} and elucidated theoretically by Eliashberg~\cite{Ivlev.1971,Ivlev.1973}, with later refinements by others~\cite{Chang.1977,Kaplan.1976,Klapwijk.1977,Tikhonov.2018fpa,Schmid.1977,Curtis.2019,Catelani.2010,Derendorf.2024,Galaktionov.1997}.
This counter-intuitive phenomenon can be understood as radiation boiling off thermally available quasiparticles from the superconducting gap edge, driving them to higher energies where their effect on superconducting order is less detrimental. This is known to occur via radiative absorption of microwaves determined by the Fermi velocity, $v_F$ of the electrons. It is most pronounced near the critical temperature $T_c$, given the near phase transition sensitivity and abundant availability of quasiparticles.

\begin{figure}
    \centering
    \includegraphics[width=1\linewidth]{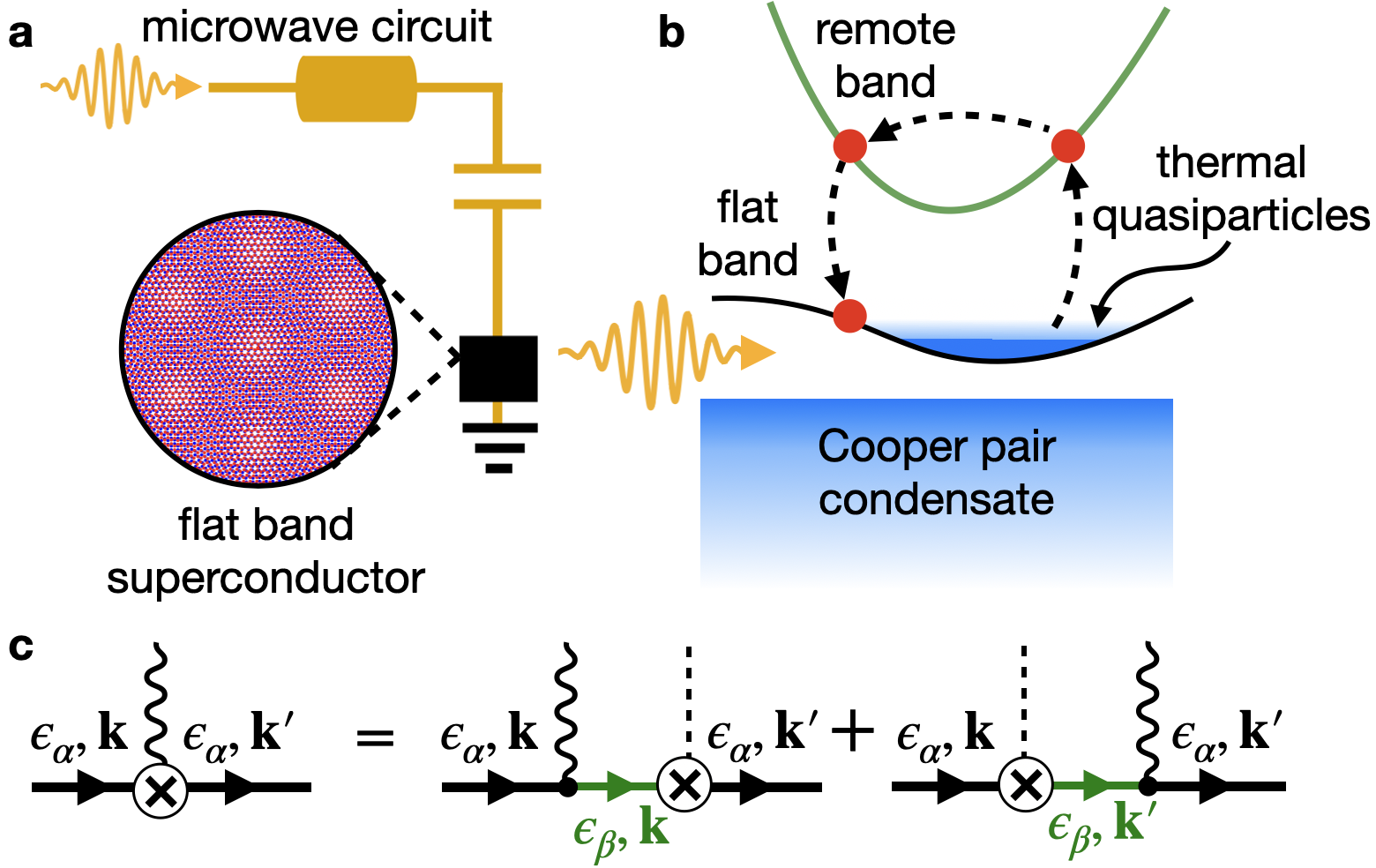}
    \caption{Microwave enhancement of superconductivity in flat band superconductors governed by quantum geometry. 
    (a) A microwave circuit with flat band superconductor for which we have picked TBG. 
    (b) When the superconductor is driven at subgap frequencies the quasiparticle spectrum is cooled-off at band bottom as thermally available quasiparticles are excited to higher energies via inelastic radiative scattering where momentum exhange is provided by disorder. In particular for flat bands, the nonequilibrium excitations are achieved via virtual transitions to proximity bands which are captured in quantum geometry. (c) The quantum geometric origin of nonequilibrium dyanmics in flat band superconductors is revealed by disorder corrected velocity vertex from where we can obtain the absorption determined by weighted interband quantum connection, see Eq~(\ref{eq:FGR-impurity}). 
    }
    \label{fig:schematic}
\end{figure}

In this {\it Letter} we will extend this idea to flat band superconductors and, in particular demonstrate a novel route towards nonequilibrium superconductivity in which quantum geometric effects play a central role.
Our key result is that despite vanishing $v_F$ in flat bands, an effective microwave absorption channel in flat bands is enabled by virtual transitions to proximal bands due to cooperative effect of quantum geometry and disorder, see Fig.~\ref{fig:schematic}. 
We demonstrate this inelastic radiative scattering and corresponding microwave enhancement of superconductivity in 1D toy model, and twisted bilayer graphene (TBG), a promising candidate material.
In TBG, we find that a large superconducting gap enhancement of nearly 20\% near $T_c$ may be achievable even at moderate drive strengths due to large microwave-frequency dynamical phase space and pronounced quantum geometry in moir\'e flat bands.
We anticipate such nonequilibrium properties could be readily observed with state-of-the-art microwave circuits, as developed in Ref.~\onlinecite{nick2024cqed} and recently applied to twisted graphene multilayers~\cite{joel2024tbgsc,cqed2024philipkim}.

\textit{Nonequilibrium flat band superconductivity}\textemdash 
To describe nonequilibrium superconductivity in flat bands, we start by considering a general multiband system which hosts at least one flat Bloch band. The corresponding Hamiltonian satisfies $\hat{\mathcal{H}}_{\vec k}\psi_{\alpha \vec{k}} = \epsilon_{\alpha\vec{k}} \psi_{\alpha \vec{k}}$ for Bloch state $\psi_{\alpha\vec k}$ at energy $\epsilon_{\alpha\vec k}$. The superconducting order can be described within the mean-field picture. Assuming $SU(2)$ spin-rotation and time reversal symmetries, we write the Bogoliubov-de Gennes (BdG) Hamiltonian
\begin{equation}
\label{eq:bdgH}
    {\check{H}}^{\rm BdG}_{\vec k} = \begin{pmatrix}
        \mathcal{\hat{H}}_{\vec k}-\mu & \hat{\Delta}_{\vec k} \\
        \hat{\Delta}^\dag_{\vec k} & -\mathcal{\hat{H}}^T_{-\vec k}+\mu
    \end{pmatrix}.
\end{equation}
where $\mu$ is the chemical potential and $\hat{\Delta}_{\vec k}$ is the superconducting gap. Here we distinguish between matrices in Bloch (indicated by a hat $\hat{\cdot }$) and Nambu (indicated by a check $\check{\cdot}$) space, and use the units where $\hbar=k_b=c=e=1$. For brevity, we assume $s$-wave pairing trivial in orbital and band basis, i.e. $\hat{\Delta}_{\vec k} = \Delta \hat{1}$. This implies $[\hat{\Delta}_{\vec k}, \hat{\mathcal{H}}_{\vec k}]=0$. 
In this case, ${\check{H}}^{\rm BdG}_{\vec k}$ can diagonalized for each band separately to give 
$ {\check{H}}_{\mathbf{k}}^{\rm BdG}\chi_{\pm \alpha \bf k}  = \pm E_{\alpha\bf k}\chi_{\alpha \pm \bf k}.
$
The energy eigenvalues are characterized by $E_{\alpha\bf k} = \sqrt{(\epsilon_{\alpha\bf k}-\mu)^2 + |\Delta|^2}$, and the Nambu spinor is written as $\chi_{\alpha+\bf k} = (u_{\alpha \bf k}, v_{\alpha\bf k})^T \otimes \psi_{\alpha \bf k}$, whereas $\chi_{\alpha -\bf k} = i\check{\tau}_2 \chi_{\alpha -\bf k}$. Here, $\check{\tau}_a$ is the Pauli matrix in Nambu space and $u_{\alpha\vec k},v_{\alpha \vec k} = [\{1\pm(\epsilon_{\alpha \vec k}-\mu)/E_{\alpha\vec k}\}/2]^{1/2}$. We fix $\mu$ in flat band $\alpha$, and the superconducting gap can be determined self-consistently using the gap equation 
\begin{equation}
\label{eq:gap}
    \frac{\Delta}{g} = \sum_{\vec k} \frac{\Delta}{2E_{\alpha\vec k}}\left[1 - 2 f_{\alpha\bf k}\right],
\end{equation}
where $g$ is the coupling constant and 
$f_{\alpha\vec k}$ is the distribution of quasiparticles which is assumed the same for both spin-species with energy $E_{\alpha\vec k}$; $\beta = 1/T$ with $T$ being the temperature. 

In presence of microwave drive, quasiparticles absorb energy and redistribute attaining a nonequilibrium distribution. From quasiparticle density of states $\sim E/\sqrt{E^2 -\Delta^2}$ and Eq.~(\ref{eq:gap}), one can readily see if the quasipartilce distribution decreases around the quasiparticle band edge, the superconducting gap increases. This is particularly pronounced near $T_c$, and at subgap frequencies $(\omega < 2\Delta)$ where disruption of superconducitivity by Cooper pair breaking is avoided. 

We determine the nonequilibrium quasiparticle distribution perturbatively such that at leading order $f_{\alpha\vec k}\approx f_{\alpha\vec k}^0 + \delta f_{\alpha\vec k}$
where $f^{0}_{\alpha\vec k} = (1+e^{\beta E_{\alpha\vec k}})^{-1}$ denotes the equilibrium distribution. Here, $\delta f_{\alpha\vec k}$ captures the effect of microwave driving, and we obtain it by solving kinetic equation in steady state~\cite{Curtis.2019}. This gives $\delta f_{\alpha\vec k} = \tau_{\rm in}\mathcal{I}_{\vec k}^{\rm rad}$ which for $\omega <2\Delta$ is determined by the inelastic radiative scattering from $\chi_{\alpha\vec k}$ to $\chi_{\alpha\vec k'}$ with transition rate $W_{\alpha\vec k \rightarrow \alpha\vec k'}$, captured by the collision integral 
\begin{equation}
\label{eq:collision_integral}
    \mathcal{I}^{\rm rad}_{\vec k} = \sum_{\bf k'} W_{\alpha {\bf k} \to \alpha \bf k'} \left[ f_{\alpha \bf k'}^{0} - f_{\alpha \bf k}^{0}\right].
\end{equation}
Additionally, $\delta f_{\alpha \vec k}$ depends on inelastic relaxation time of quasiparticles, $\tau_{\rm in}$. 
We use a phenomenological $\tau_{\rm in}$ in this work, however, often this is determined by interactions of electrons with phonon heat bath~\cite{Chang.1977,Kaplan.1976}.

{\it Superconducting gap enhancement}\textemdash 
We solve for inelastic radiative scattering in Eq.~(\ref{eq:collision_integral}), and demonstrate how it enables microwave enhancement of superconducting gap. However, momentum transfer from microwave radiation is negligible and $\mathcal{I}_{\vec k}^{\rm rad}$ for scattering in the same quasiparticle band would vanish due to momentum conservation. Nevertheless, the momentum conservation can be relaxed due to disorder which is known to be crucial in describing electromagnetic absorption in single-band superconductors~\cite{Mattis.1958}. Thus, it is important to consider the combined effect of microwave drive and disorder.   

We first consider the effect of microwave radiation which is incorporated by Peierl's substitution,  $\mathbf{k} \to \mathbf{k} - {\mathbf{A}}_{\rm ext}\check{\tau}_3$ for a homogeneous electric field described by vector potential $\mathbf{A}_{\rm ext}$~\footnote{We choose Weyl gauge with scalar potential $\phi = 0 $ such that $\mathbf{E} = -\partial_t \mathbf{A}_{\rm ext}$}. 
The Hamiltonian that describes the interaction between microwave radiation and quasiparticles, to linear order in field, is thus
\begin{equation}
\label{eq:radiation}
    \check{H}^{\rm rad}_{\bf k} = -\begin{pmatrix}
        \partial_{\vec k} \mathcal{\hat{H}}_{\vec k} & 0 \\
        0 & \partial_{\vec k} \mathcal{\hat{H}}^T_{-\vec k} \\
    \end{pmatrix}\cdot\mathbf{A}_{\rm ext} \equiv - \mathbf{\hat{v}}_{\bf k} \cdot\mathbf{A}_{\rm ext}\check{\tau}_0
\end{equation}
where we have introduced the Bloch velocity operator $\hat{\vec v}_{\vec k} = \partial_{\vec k}\hat{\mathcal{H}}_{\vec k}$. Typically, this interaction in the single band limit describes conventional radiation-stimulated superconductivity where the absorption is estimated to go as $\sim|v_F|^2$. 
This conventional paradigm fails for flat bands as $v_F \rightarrow 0$, quenching microwave absorption in flat band superconductors. 
In a multiband system however, this problem is circumvented due to the presence of nontrivial band geometry which gives contributions to the velocity vertex off-diagonal in band-space, as $[ {\check{H}}_{\bf k}^{\rm rad}, \check{H}^{\rm BdG}_{\bf k}] \neq 0$ in general. Given the pronounced role of quantum geometry in flat band superconductivity, we systematically utilize this in our treatment below. 

Next, the disorder is accounted by short-range random impurities $\hat{V}^{\rm imp}(\mathbf{r})= \sum_j V_j \delta(\vec r - \vec R_j)$ modelled with a Gaussian-distributed field $V_j$ which satisfies $\langle V_j^2\rangle_{\rm dis} = N_{\rm imp} V_0^2$~\cite{mahan2013textbook}. Here, $N_{\rm imp}$ is the impurity concenteration and $V_0$ denotes the strength of disorder. 
The disorder then enters in to the BdG Hamiltonian as $\check{{H}}^{\rm BdG}_{\bf k',k} \to \delta_{\bf k',k}\check{H}^{\rm BdG}_{\bf k} + \check{\tau_3} \hat{V}^{\rm imp}_{\bf k',k}$.

Finally, we proceed to evaluate the transition rate $W_{\alpha\vec k \rightarrow \alpha\vec k'}$ which can be computed using the Fermi's Golden Rule as
\begin{equation}
\label{eq:goldenrule}
    W_{\alpha\vec k \rightarrow \alpha\vec k'} = 2\pi |\mathbf{A}_{\rm ext}|^2 |\chi_{\alpha \bf k'}^\dagger \hat{\bm\Gamma}_{\bf k k'} \check{\tau}_0 \chi_{\alpha \bf k }|^2     \delta(E_{\alpha\vec k'} - E_{\alpha\vec k} - \omega).
\end{equation}
We restrict our treatment to processes which remain in the same band around the Fermi surface $\alpha$ for simplicity. The effective velocity vertex $\hat{\bm\Gamma}_{\bf k k'}$ differs from $\hat{\vec v}_{\vec k}$, see Eq.~(\ref{eq:radiation}), in the presence of disorder. Here, $\hat{\bm\Gamma}_{\bf k'k}$ is obtained by expanding the Bloch states up to leading order in impurity potential. This yields
\begin{equation}
\label{eq:effective-vertex}
  \hat{\bm\Gamma}_{\bf k k'} = \sum_{\beta \neq \alpha}\hat{\mathbf{v}}^{\alpha \beta}_{\bf k}\frac{\psi^\dagger_{\beta \bf k} \hat{V}^{\rm imp}_{\bf k',k} \psi_{\alpha \bf k'} }{\epsilon_{\beta\bf k} - \epsilon_{\alpha \bf k'}} +  \frac{\psi^\dagger_{\alpha \bf k} \hat{V}^{\rm imp}_{\bf k',k} \psi_{\beta \bf k'} }{\epsilon_{\beta \bf k'} - \epsilon_{\alpha \bf k}}\hat{\mathbf{v}}^{\beta \alpha}_{\bf k'},
\end{equation}
which systematically captures the disorder assisted inelastic radiative scattering in a flat band. Importantly, Eq.~(\ref{eq:effective-vertex}) reveals virtual transitions to proximal bands highlighting the role of Bloch quantum geometry in multiband systems which is particularly relevant in the case of flat bands, illustrated in Fig.~\ref{fig:schematic}b,c. The effective velocity vertex can also be obtained by a Schrieffer-Wolff procedure, as for instance recently developed in Ref.~\onlinecite{Mao.2023,Mao.2024}; details are relegated to the Supplemental Material (SM). We simplify Eq.~(\ref{eq:effective-vertex}) by utilizing $\check{\tau}_0$ Nambu channel for radiative scattering, and after disorder average
\begin{multline}
\label{eq:FGR-impurity}
  W_{\alpha\vec k \rightarrow \alpha\vec k'} = 2\pi N_{\rm imp} V_0^2 |\mathbf{A}_{\rm ext}|^2 L_{\vec k \vec k'} \\
  \Phi_{\bf k k'} \delta(E_{\alpha\vec k'} - E_{\alpha\vec k} - \Omega) 
\end{multline}
where $L_{\vec k\vec k'} = (u_{\alpha\vec k}u_{\alpha\vec k'} + v_{\alpha\vec k}v_{\alpha\vec k'})^2$ is given by superconducting coherence factors determining probability of transition based on energy of quasiparticles. Strikingly, 
\begin{equation}
    \Phi_{\vec k\vec k'} = \bigg|\sum_{\beta\neq \alpha} i\mathcal{A}^{\alpha\beta}_{\vec k\vec k'} +( {\rm h.c.},\vec k \leftrightarrow \vec k')\bigg|^2
\end{equation}
is determined completely by the quantum geometry of Bloch bands captured in weighted interband Berry connection, $\mathcal{A}^{\alpha\beta}_{\vec k\vec k'} = -i\hat{\vec v}_{\vec k}^{\alpha\beta}\psi^\dagger_{\beta \bf k} \psi_{\alpha \bf k'} / (\epsilon_{\beta\bf k} - \epsilon_{\alpha \bf k'})$, modified by the band overlap $\psi^\dagger_{\beta \mathbf{k}}\psi_{\alpha\bf k'}$ due to the scattering by disorder.
As such, this vanishes as $\mathbf{k}\to \mathbf{k}'$.
Interestingly, we find that the presence of proximal bands prevents the quenching of quasiparticle dynamics even though $v^{\alpha \alpha}_{k} \sim v_F\rightarrow 0$ in flat bands and enables microwave-enhanced flat band conductivity.

The transition rate in Eq.~(\ref{eq:FGR-impurity}) along with $\mathcal{I}^{\rm rad}_{\vec k}$ in Eq.~(\ref{eq:collision_integral}) can be used to obtain the change in quasiparticle distribution, $\delta f_{\vec k}$, given which one can compute the change in the superconducting gap. We obtain the change in superconducting gap in the Ginzburg-Landau regime, $T\lesssim T_c$. For a flat band of bandwidth $W$
\begin{equation}
\label{eq:changegap}
    \frac{\delta\Delta}{\Delta_0} \approx a_0^{-1} \sum_{\vec k} \frac{\delta f_{\alpha\vec k}}{E_{\alpha\vec k}}
\end{equation}
where $\Delta_0$ is the superconducting in absence of drive, $a_0 = \nu_F [(T-T_c)/T_c]\tanh(\beta_c W/4)$ in the limit $g W \lesssim 1$, and $\nu_F$ is the flat band single particle density of states near the Fermi energy. The details to obtain Eq.~(\ref{eq:changegap}) are presented in SM.

\begin{figure}
    \centering
    \includegraphics[width=1\linewidth]{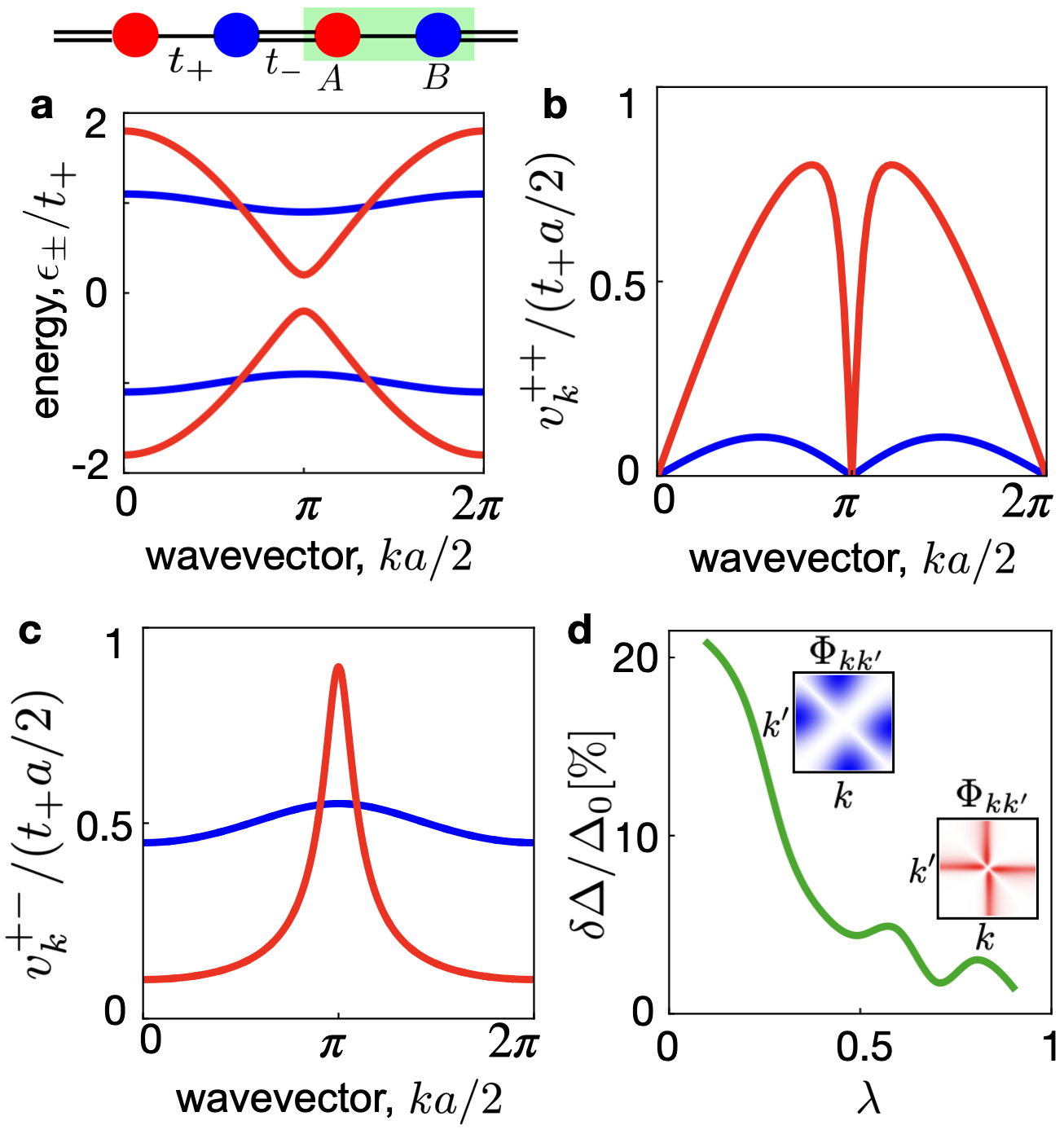}
    \caption{Microwave enhancement of superconducting gap in 1D SSH model. (a) 1D bipartite lattice with nearest neighbor intra- and inter-cell hoppings denoted by $u$ and $v$ respectively. The unit cell is shown as the green box. The model has bandwidth tunable by $\lambda = t_-/t_+$. Here, we show the band structure for $\lambda = 0.1$ (blue) and 0.8 (red). (b,c) Distribution of intra- and interband  velocity matrix elements, $v^{+\pm}_k$, for values of $\lambda$ corresponding to dispersion in panel (a). Reduced bandwidth implies suppressed $v^{++}_k$. On the other hand, for narrow bands large $v^{+-}_k$ is distributed throughout the $k$-space. (d) Quantum geometry governed superconducting gap enhancement in the 1D model obtained from Eq~(\ref{eq:changegap}). Evidently, $\delta\Delta/\Delta_0$ decreases with increasing $\lambda$ (bandwidth) highlighting the role of pronounced interband coherences in flat band systems. The inset shows distribution of $\Phi_{kk'}$ color coded with dispersion in panel(a). The spread of $v^{+-}_k$ is mimicked in the distribution of $\Phi_{kk'}$. Here we use $\mu = u$, $\Delta=0.1 u$,  $\omega=0.005 u$, $T/T_c = 0.1$ and set $V_0^2 N_{\rm imp} e^2 |\mathcal{E}|^2/(\hbar\tau^{-1}) = 0.3 u^3a^{-2}$.}
    \label{fig2}
\end{figure}

{\it Gap enhancement in 1D topoogical bands}\textemdash 
To elucidate the physical nature of quantum geometry enabled microwave-enhanced flat band superconductivity, we consider a 1D Su-Schreiffer-Heeger (SSH) model which has local singlet $s$-wave pairing. 
This is depicted in Fig. \ref{fig2}a, where we consider a 1D bipartite lattice with nearest neighbor intra- and inter-cell hopping, $t_+$ and $t_-$, respectively. 
In the ordered basis $\{|k,A\rangle, |k,B\rangle\}$, the systems's Bloch Hamiltonian is
\begin{equation}
    \label{eq:1dmodel}
    \hat{\mathcal{H}}_{\bf k}^{\rm 1D} = \begin{pmatrix}
        0 & J_{k}\\
        J_k^* & 0
    \end{pmatrix}
\end{equation}
where $J_{k} = t_+ e^{ika/2} + t_- e^{-ika/2}$ and $a$ is the lattice constant. The key feature of the Hamiltonian in Eq.~(\ref{eq:1dmodel}) is the tunable bandwidth with parameter $\lambda = t_-/t_+$. The band structure for different values of $\lambda$ is shown in Fig. \ref{fig2}a. Evidently, $\lambda\rightarrow 0$ denotes the flat band condition. Note that intraband, $v^{++}_k = -2 \lambda at_+^2  \sin ka/|J_k|$, and interband, $v^{+-}_k = (1-\lambda^2) at_+^2/|J_k|$, velocity matrix elements are vanish for $\lambda=0,1$ respectively; we have used $\pm$ to denote bands with energy $\epsilon_{\pm} = \pm |J_k|$. 
The distribution of $v^{++}_k$ and $v^{+-}_k$ in the Brillouin zone (BZ) is shown in Fig. \ref{fig2}b and \ref{fig2}c, respectively. We note large interband velocity matrix element distributed throughout the BZ for flat band which is in contrast to dispersive bands where interband velocity matrix element is concenterated only around the band edge. Finally, we obtain the the change in superconducting order using Eq.~(\ref{eq:changegap}) in variation with $\lambda$. Evidently, $\delta\Delta/\Delta_0$ decreases with increasing $\lambda$ which highlights the striking dependence of nonequilibrium many body dynamics on quantum geometry in flat bands.

{\it Gap enhancement in TBG}\textemdash We now examine the microwave-enhanced superconductivity in a promising candidate system: TBG. 
It is a particularly attractive venue for microwave enhanced superconductivity due to the flat bands and strong interactions, with typical values of the interaction strength that can even exceed the bandwidth of the narrow bands. Importantly, near magic angle, superconductivity with $T_c\approx 3$K emerges around half-filling and its quantum geometric origin has been verified in experiments~\cite{lau2023tbgsc,joel2024tbgsc}. 

\begin{figure}
    \centering
    \includegraphics[width=1\linewidth]{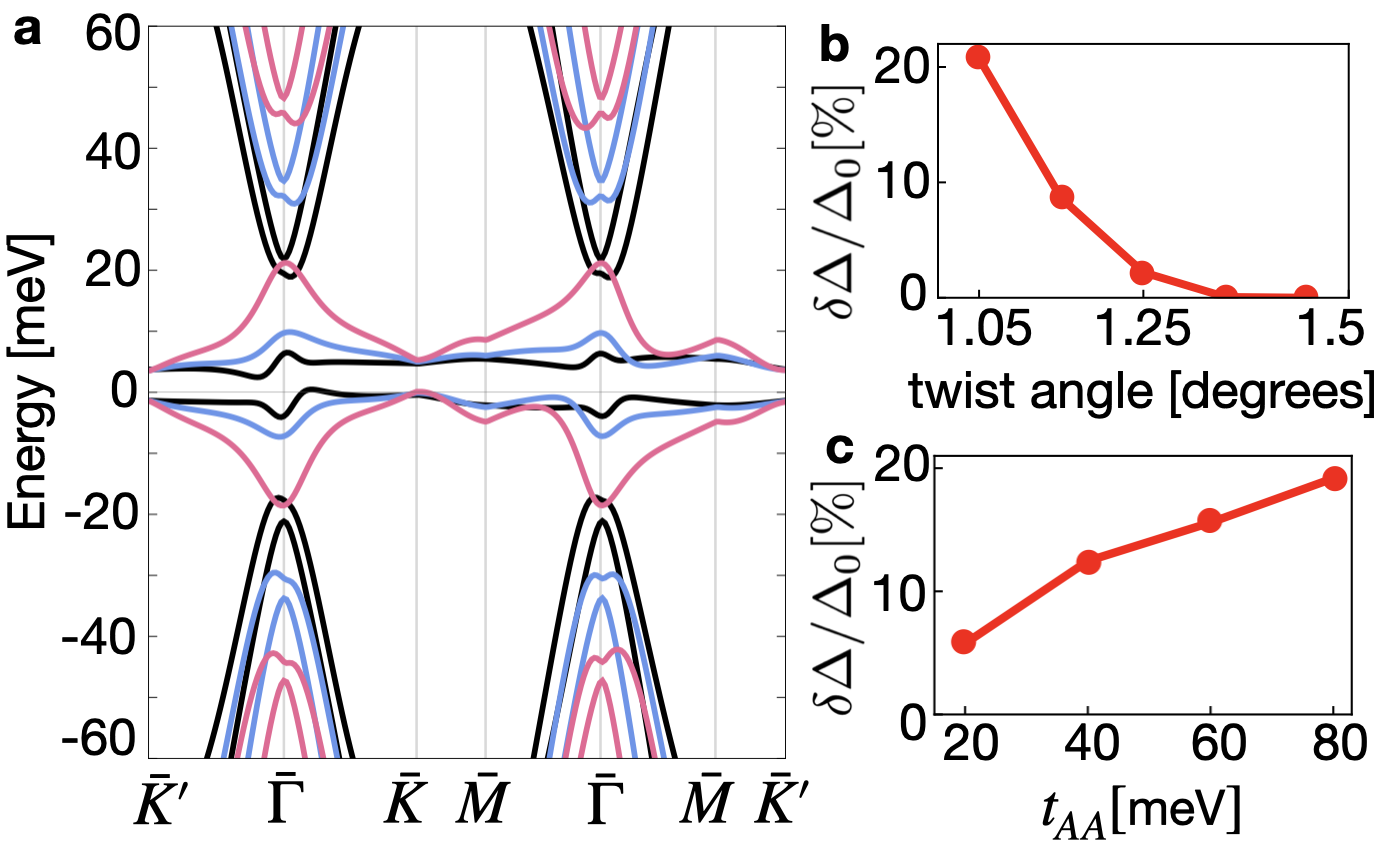}
    \caption{Microwave enhancement of superconducting gap in TBG. (a) TBG band structure in the moire BZ at various twist angle $\theta=1.05^\circ$ (black), $1.15^\circ$ (blue) and $1.25^\circ$ (pink). $\theta=1.05^{\circ}$ is the magic angle with minimum band width. Here, we include lattice relaxation, hBN encapsulation and strain, and use following parameters (see SM for details): $t_{AA} = 79.7$ meV, $t_{AB} = 97.5$ meV, $\delta_1 = \delta_2 = 5$ meV, strain = 0.1\%. (b) Variation of $\delta\Delta/\Delta_0$ with twist angle which is most pronounced at magic angle. We use same parameters as in (a), and chemical potential, $\mu = 3$ meV, inelastic relaxation time $\tau_{\rm in} = 1$ ps, $N_{\rm imp}V_0^2 = 100$ meV$^2$, $T/T_c=0.1$, $\Delta_0 = 0.1$ meV, $\omega=0.05$ meV and $|\vec{E}| = 20$ V/cm. (c) $\delta\Delta/\Delta_0$ in variation with $t_{AA}$ which elucidates the effect of proximity of remote bands. The gap between remote and flat bands is tunable by the relative magnitudes of $t_{AA}$ and $t_{AB}$. Here we fix $t_{AB}$ as in (a) at magic angle, chemical potential at $\mu = 5$ meV, $|\vec{E}| = 10$ V/cm and vary $t_{AA}$. }
    \label{fig3}
\end{figure}

Guided by this we focus on nonequilibrium superconductivity in near-magic angle TBG around half-filling. We model the electrons in this system using the continuum model, and include modest heterostrain and the effects from hBN encapsulation of the graphene layers \cite{arora2022quantum}. This produces a set of narrow bands tunable by twist angle, see Fig.~\ref{fig3}a.  
For convenience of the reader, details of the Hamiltonian are included in SM. For superconducting order, we consider $s$-wave pairing which has been previously used for a qualitative description of TBG superconductivity~\cite{bernevig2019-tbgsc-phonon,torma2020-tbgsc,ktlaw2024ginzburg}. 

We first analyze the effect of bandwidth on quantum geometry determined $\delta\Delta/\Delta_0$ in TBG heterostructures. While superconductivity is found only around the magic angle, the twist angle can be used as a knob to tune the bandwidth for illustration. We set the chemical potential in the conduction flat band and numerically investigate the gap enhancement using Eq.~(\ref{eq:changegap}) in variation with the twist, see SM for details. Strikingly, $\delta\Delta/\Delta_0$ is maximum at $\theta = 1.05^{\circ}$ where the bandwidth is minimum and we obtain large $\delta\Delta/\Delta_0$ of about 20\% for a modest drive strength, see Fig. \ref{fig3}a,b. This is consistent with pronounced optoelectronic THz responses governed by quantum geometry at magic angle~\cite{arora2021strain,chaudhary2022shift,Kumar.2024}. In these calculations, we have modelled the inelastic relaxation of excited quasiparticles by using typical electron-phonon scattering rates in graphene heterostructures ($\tau_{\rm in} = 1$ps~\cite{wu2019phonon}). 

Next, we investigate the role of proximal remote bands on the nonequilbrium superconductivity. This is particularly relevant for TBG heterostructures where lab procedures and interactions have been known to cause macroscopic differences in physical properties \cite{lau2022reproducibility, uri2020mapping, hesp2021observation}. At a qualitative level, this can be accounted with lattice relaxation effects which can alter the separation between flat and remote bands \cite{cscmtbg,namkoshino}. Within the continuum Hamiltonian lattice relaxation is parameterized by the ratio between interlayer hopping amplitudes between same and different lattice sites, i.e., $t_{AA}/t_{AB}$.  In Fig. \ref{fig3}c, we fix $t_{AB}$ and show the gap enhancement at different values of $t_{AA}$. Note that decreasing $t_{AA}/t_{AB}$ reduces the bandwidth of flat bands and pushes remote bands higher in energy  
(see SM for TBG band structure for different values of $t_{AA}/t_{AB}$). Since, the coherence between energetically far bands is suppressed, we note reduced $\delta\Delta/\Delta_0$ for smaller values of $t_{AA}$ revealing the role of proximal remote bands in quantum geometry enabled nonequilibrium superconductivity in TBG. 

{\it Conclusion}\textemdash 
In this work we have shown that the quantum geometry of Bloch bands in a flat-band superconductor can enable a novel mechanism for inducing nonequilibrium dynamics, with the potential for dramatic microwave-enhancement of the superconducting gap. 
Our work underscores the pivotal role of nontrivial Bloch wavefunction winding in flat bands, while also highlighting the role of remote bands, which are particularly relevant in moir\'e heterostructures, such as TBG. While we have concentrated on microwave driving by ``classical" light, we emphasize that the features of nonequilibrium flat band superconductivity proposed here are likely also relevant to cavity systems wherein the cavity may enable additional control via the quantum nature of light~\cite{Curtis.2019}. 
Furthermore, the tunablity of correlated phases in flat band systems suggests that such nonequilibrium functionalities can be extended to other correlated phases, e.g., spontanously generated fractional and integer anomalous Hall excitations in graphene multilayers~\cite{lu2024fractional,choi2024electric}. 
It is known that microwave driving may have a dramatic effect on the nonequilibrium physics of Landau levels in magnetic field~\cite{Mani.2002,Mani.2004,Zudov.2001,Dmitriev.2003,Dmitriev.2004,Dmitriev.2008,Dmitriev.2012,Vavilov.2004,Auerbach.2005,Durst.2003}, and it could potentially be interesting to extend this to strongly interacting Chern bands in moir{\'e} materials. Additionally, details involving electron-phonon dynamics~\cite{Knap.2016,Babadi.2017,Murakami.2017}, strongly-coupled superconducting channels~\cite{strongSC,strongphasesTBG,strongTBGexperiment} and interaction modified flat bands~\cite{cea2020band} are open directions for future works. 
From a technological perspective, this can be used for better control and accessibility of the superconducting phase which can be utilized in superconductor devices for sensing of quantum matter.

{\it Acknowledgements}\textemdash 
The authors acknowledge useful discussions with Amir Yacoby, Gil Refael, Justin Song, Sayed Ali Akbar Ghorashi, Nicholas R. Poniatowski, Marco Polini and Joel Wang. This work was supported by Quantum Science Center, a National Quantum Information Science Center of the U. S. Department of Energy, Gordon and Betty Moore Foundation Grant No. 8048, and the John Simon Guggenheim Memorial Foundation (Guggenheim Fellowship).
This work was performed in part at Aspen Center for Physics, which is supported by National Science Foundation grant PHY-2210452.
JBC acknowledges support from ARO grant number W911NF-21-1-0184 and the SNSF project 200021\_212899.

\bibliographystyle{apsrev4-1}
\bibliography{nq_flat_band_sc_arXiv_upload}

\appendix
\onecolumngrid
\newpage
\pagebreak
\widetext
\setcounter{equation}{0}
\setcounter{figure}{0}
\setcounter{table}{0}
\setcounter{page}{1}
\makeatletter
\renewcommand{\theequation}{S\arabic{equation}}
\renewcommand{\thefigure}{S\arabic{figure}}
\renewcommand{\bibnumfmt}[1]{[S#1]}

\section{Supplemental Material for ``Quantum geometry induced microwave enhancement of flat-band superconductivity''}

\subsection{Adiabatic Elimination of Remote Bands}

Here we provide the alternative, and equivalent, procedure for obtaining the effective velocity operator $\check{\Gamma}$.
We closely follow Refs.~\onlinecite{Mao.2023,Mao.2024}.
We first consider the normal state with $\Delta = 0$ though we retain the Nambu space, so that we have full second-quantized BdG Hamiltonian of 
\begin{equation}
    {H} = \sum_{\bf k,k'} \Psi_{\bf k'}^\dagger \left[ \check{\tau}_3 \hat{\mathcal{H}}_{{\bf k} - \mathbf{A}(t)\check{\tau}_3}\delta_{\bf k',k}  + \check{\tau}_3\hat{V}_{\bf k',k}^{\rm imp}\right]\Psi_{\bf k}.
\end{equation}
Here $\Psi_{\bf k}$ is the Nambu annihilation operator for electrons in the multiband basis. 
Here we have used the Peierl's substitution on the normal-state Bloch Hamiltonian to obtain the effect of the gauge potential, and $\hat{V}_{\bf k',k}^{\rm imp}$ is the impurity scattering potential and is assumed to respect time-reversal symmetry.
Even if this is a trivial identity matrix in the orbital basis, such that $\hat{V}_{\bf k',k}^{\rm imp} = V_{\bf k',k}^{\rm imp}\hat{1}$, this will still obtain nontrivial overlaps between Bloch bands of the form 
\begin{equation}
    U^{\beta\alpha}_{\bf k'k} \equiv \langle \beta \mathbf{k}' |\alpha  \mathbf{k}\rangle  V_{\bf k',k}^{\rm imp}.
\end{equation}
The Bloch bands are denoted by $|\alpha \mathbf{k}\rangle$ with Bloch band eigenvalue $\epsilon_{\alpha \bf k}$. 

It is useful to partition these bands in to two sets; the ``flat bands" with energies $|\epsilon_{\alpha \bf k} | < \epsilon_c$ within a cutoff window of the Fermi level, and ``remote bands with energies $|\epsilon_{\alpha \bf k} | > \epsilon_c$ outside this cutoff window.
Presumably $\epsilon_c$ is large as compared to the energy scales of the flat-band band-width or superconductivity. 

To this end, we introduce the projection operators, defined for every $\mathbf{k}$ point separately, as 
\begin{equation}
    \mathcal{\hat{P}}_{\bf k} = \sum_{\alpha} |\alpha\mathbf{k}\rangle \langle \alpha \mathbf{k}| \theta( |\epsilon_{\alpha\bf k}| < \epsilon_c ) ,
\end{equation}
and its complement $\mathcal{\hat{Q}}_{\bf k} = \hat{1} - \mathcal{\hat{P}}_{\bf k}$. 
This involves the logical indicator function $\theta(x)$ which is one if $x$ is true, and $0$ if $x$ is false. 

Now, we perform Schrieffer-Wolff elimination of the electrons in these remote bands. 
For the linear case at hand, this is easiest done by taking the Heisenberg equations of motion. 
The equations of motion for the Nambu electrons is obtained as 
\begin{equation}
\left[ i\partial_t - \hat{\mathcal{H}}_{\bf k}\check{\tau}_3 \right] \psi_{\bf k} = -\mathbf{A}(t)\cdot \check{\mathbf{v}}_{\bf k} \psi_{\bf k} + \hat{V}^{\rm imp}_{\bf k,k'}\check{\tau}_3 \psi_{\bf k'} .
\end{equation}
Here we have gone to the Bloch band basis, with $\psi_{ \bf k}$ the Nambu-Bloch spinor for electrons.
We have also expanded up to linear order in the perturbing electromagnetic potential, and defined the velocity operator
\begin{equation}
    \check{\bf v}_{\bf k}  =  \check{\tau}_0\partial_{\bf k}\hat{\mathcal{H}}_{\bf k} . 
\end{equation}
This has matrix elements 
\begin{equation}
    \langle \alpha\mathbf{k}|\check{\bf v}_{\bf k} |\beta\mathbf{k}\rangle = \mathbf{v}^{\alpha\beta}_{\bf k} . 
\end{equation}

We can now solve this equation by solving first for the remote bands; if we are only interested in low-energy properties, we can neglect $i\partial_t \ll \epsilon_c$, so that 
\begin{equation}
 -  \mathcal{\hat{Q}}_{\bf k} \mathcal{\hat{H}}_{\bf k}\check{\tau}_3 \psi_{\bf k} = -\mathcal{\hat{Q}}_{\bf k}\mathbf{A}(t)\cdot\check{\mathbf{v}}_{\bf k}\mathcal{\hat{P}}_{\bf k} \psi_{\bf k} + \sum_{\bf k'} \mathcal{\hat{Q}}_{\bf k}\hat{V}^{\rm imp}_{\bf k,k'}\check{\tau}_3 \mathcal{\hat{P}}_{\bf k'}\psi_{\bf k'} .
\end{equation}
We have neglected the terms diagonal in $\mathcal{\hat{Q}} \cdot \mathcal{\hat{Q}}$ since we assume the bands are intrinsically unoccupied as $\epsilon_c \gg T$, so that $\hat{\mathcal{Q}}\psi$ can be regarded as near zero, and all occupation of the remote bands comes from the lower bands virtually scattering in to and out of them.
We can then solve this and insert it in to equations of motion for the flat bands.
Note $[\mathcal{\hat{Q}}_{\bf k}, \hat{\mathcal{H}}_{\bf k}] = 0$. 

This is split between intra-flat band terms and terms which arise from perturbative dressing by the remote bands at second order, with  
\begin{multline}
\left[ i\partial_t -  \hat{\mathcal{H}}_{\bf k}\check{\tau}_3 \right] \mathcal{\hat{P}}_{\bf k}\psi_{\bf k} =  -\mathcal{\hat{P}}_{\bf k} \mathbf{A}(t)\cdot \check{\mathbf{v}}_{\bf k}\mathcal{\hat{P}}_{\bf k} \psi_{\bf k} + \sum_{\bf k'} \mathcal{\hat{P}}_{\bf k} \hat{V}^{\rm imp}_{\bf k,k'}\check{\tau}_3 \mathcal{\hat{P}}_{\bf k'}\psi_{\bf k'}  \\
 - \mathbf{A}(t)\cdot\mathcal{\hat{P}}_{\bf k}\check{\mathbf{v}}_{\bf k}\mathcal{\hat{Q}}_{\bf k}  \left(-  \check{\tau}_3\mathcal{\hat{Q}}_{\bf k} \mathcal{\hat{H}}_{\bf k} \mathcal{\hat{Q}}_{\bf k}\right)^{-1} \left[ -\mathbf{A}(t)\cdot\mathcal{\hat{Q}}_{\bf k} \check{\mathbf{v}}_{\bf k} \mathcal{\hat{P}}_{\bf k} \psi_{ \bf k} + \mathcal{\hat{Q}}_{\bf k} \sum_{\bf k'}\hat{V}^{\rm imp}_{\bf k,k'} \check{\tau}_3 \mathcal{\hat{P}}_{\bf k'} \psi_{\bf k'}\right]\\
 + \sum_{\mathbf{k}'}  \mathcal{\hat{P}}_{\bf k} \hat{V}_{\rm k,k'}^{\rm imp}\check{\tau}_3  \mathcal{\hat{Q}}_{\bf k'}\left(-  \check{\tau}_3\mathcal{\hat{Q}}_{\bf k'} \mathcal{\hat{H}}_{\bf k'} \mathcal{\hat{Q}}_{\bf k'}\right)^{-1}\left[ -\mathbf{A}(t)\cdot\mathcal{\hat{Q}}_{\bf k'} \check{\mathbf{v}}_{\bf k'} \mathcal{\hat{P}}_{\bf k'} \psi_{ \bf k'} + \mathcal{\hat{Q}}_{\bf k'} \sum_{\bf k''}\hat{V}^{\rm imp}_{\bf k',k''} \check{\tau}_3 \mathcal{\hat{P}}_{\bf k''} \psi_{\bf k''}\right].
\end{multline}
Of the four terms generated by perturbation, we see the last one is second order in impurity scattering and therefore will only renormalize the already-present impurity scattering.
We therefore ignore this term as it is not qualitatively important. 
Likewise, we see that there is a term generated at second order in the velocity operator. 
While this is in principle an interesting and important term to retain for studying the nonlinear response, or computing superfluid density for instance, we see that this will not generate the relevant inelastic scattering processes at second order since it does not scatter the momentum. 

We therefore retain only the two cross terms, which we can use to define an effective velocity operator with matrix elements  
\begin{equation}
 \check{\bm\Gamma}_{\bf k\bf k'} = \mathcal{\hat{P}}_{\bf k}\check{\mathbf{v}}_{\bf k} \mathcal{\hat{Q}}_{\bf k}\left(-  \check{\tau}_3\mathcal{\hat{Q}}_{\bf k} \mathcal{\hat{H}}_{\bf k} \mathcal{\hat{Q}}_{\bf k}\right)^{-1}\mathcal{\hat{Q}}_{\bf k}\hat{V}^{\rm imp}_{\bf k,k'} \check{\tau}_3 \mathcal{\hat{P}}_{\bf k'} + \mathcal{\hat{P}}_{\bf k}\hat{V}^{\rm imp}_{\bf k,k'} \check{\tau}_3 \mathcal{\hat{Q}}_{\bf k'}\left(-  \check{\tau}_3\mathcal{\hat{Q}}_{\bf k'} \mathcal{\hat{H}}_{\bf k'} \mathcal{\hat{Q}}_{\bf k'}\right)^{-1} \mathcal{\hat{Q}}_{\bf k'}\check{\mathbf{v}}_{\bf k'} \mathcal{\hat{P}}_{\bf k'}.
\end{equation}
The explicit expression in terms of matrix elements is 
\begin{equation}
 {\bm \Gamma}^{\alpha\beta}_{\bf k,k'} \equiv \langle \alpha \mathbf{k}|\check{\bm\Gamma}_{\bf k\bf k'}|\beta \mathbf{k}'\rangle = -\sum_{\gamma:\ |\epsilon_{\gamma\mathbf{k}}| > \epsilon_c}\mathbf{v}^{\alpha\gamma}_{\bf k}\frac{1}{\epsilon_{\gamma\mathbf{k}}} U^{\gamma \beta}_{\bf k,k'} - \sum_{\gamma:\ |\epsilon_{\gamma\mathbf{k'}}| > \epsilon_c} U^{\alpha \gamma}_{\bf k,k'} \frac{1}{\epsilon_{\gamma\mathbf{k}'}}  \mathbf{v}^{\gamma\beta}_{\bf k'}.
\end{equation}
This corresponds to a coherent sum over the possible scattering pathways which couple to a remote band at second order and scatter back off of the external vector potential. 
This then renormalizes the effective Hamiltonian for the flat-bands by introducing additional couplings to the vector potential of the form 
\begin{equation}
\left[ i\partial_t - \mathcal{\hat{H}}_{\bf k}\check{\tau}_3 \right] \mathcal{\hat{P}}_{\bf k}\psi_{\alpha \bf k} = -\mathbf{A}(t) \cdot\left[ \mathcal{\hat{P}}_{\bf k}\check{\mathbf{v}}_{\bf k}\mathcal{\hat{P}}_{\bf k} \psi_{\bf k}  + \check{\bm \Gamma}_{\bf k,k'}\mathcal{\check{P}}_{\bf k'} \psi_{\bf k'}  \right] .
\end{equation}
This has the benefit of only involving the low-energy degrees of freedom in the flat-band, valid up to first order in the external vector potential. 
Crucially, the second term leads to intra-flat band scattering. 
The matrix element for scattering is then obtained in terms of the impurity potential as 
\begin{equation}
|{\bm \Gamma}^{\alpha\beta}_{\bf k\bf k'} |^2 = |V_{\bf k',k}^{\rm imp}|^2  \bigg|\sum_{\gamma:\ |\epsilon_{\gamma\mathbf{k}}| > \epsilon_c}\mathbf{v}^{\alpha\gamma}_{\bf k}\frac{1}{\epsilon_{\gamma\mathbf{k}}} \langle \gamma\mathbf{k}| \beta{\bf k'}\rangle + \sum_{\gamma:\ |\epsilon_{\gamma\mathbf{k'}}| > \epsilon_c} \langle  \alpha {\bf k}|\gamma\mathbf{k}'\rangle \frac{1}{\epsilon_{\gamma\mathbf{k}'}}  \mathbf{v}^{\gamma\beta}_{\bf k'}\bigg|^2 \equiv   N_{\rm imp} V_{0}^2 \Phi_{\bf k,k'}^{\alpha \beta}  .
\end{equation}
For short-range impurity scattering we have the impurity averaged result of $\overline{|V_{\bf k',k}^{\rm imp}|^2}  =  N_{\rm imp} V_{0}^2$ for the momentum scattering matrix elements. 
In the main text we will mostly be concerned with the case where we have only one flat band of interest at the Fermi level, in which case we simply drop the matrix elements on $\Phi$. 
Futhermore, if one derives this using standard perturbation theory, the modification is only in $-1/\epsilon_{\gamma \bf p} \to \frac{1}{\epsilon_{\mu \mathbf{p}'} - \epsilon_{\gamma \bf p}}$ where $\mu,\mathbf{p,p'}$ are as appropriate for that term in the summation, where $\mu \mathbf{p}'$ will run over the currently considered flat band.
The corrections due to this are small in $\epsilon_{\mu\mathbf{p}}/\epsilon_c$ and therefore are controlled.

\subsection{Superconducting gap enhancement in the Landau-Ginzburg regime}
In this section, we briefly describe the enhancement of superconducting order in the Landau-Ginzburg regime. We start with the self-consistent gap equation, see Eq.~(\ref{eq:gap}) in the main text, and write the superconducting gap under an external drive as $\Delta \approx \Delta _0 + \delta\Delta$. Next, we collect the terms which capture the effect of microwave radiation such that 
\begin{equation}
\label{eq:collect_changegap}
    \frac{\delta\Delta}{\Delta_0}\left(\frac{1}{g} - \sum_{\vec k} \frac{1-2 f_{\vec k}^0}{2E_{\vec k}}\right) = -\sum_{\vec k} \frac{\delta f_{\vec k}}{2E_{\vec{k}}}
\end{equation}
where as described in the main text $f_{\vec k}^0$ is the equilibrium distribution and $\delta f_{\vec k}$ is the correction under microwave drive. 

To proceed, we show that for $T\lesssim T_c$, the bracket in Eq.~(\ref{eq:collect_changegap}) is related to coeffcients $a_0$ and $b_0$ of the Ginzburg Landau free energy: $\mathcal{F}[\Delta_0] = a_0 \Delta_0^2 + b_0\Delta_0^4/2$. This can be done in 2-steps. First, we write the quasiparticle energy denominator on the left hand side of Eq.~(\ref{eq:collect_changegap}) as $E_{\vec k}^{-1}\approx |\epsilon_{\vec k} - \mu|(1-\Delta^2/2|\epsilon_{\vec k} - \mu|^2)$. Second, we compare the respective terms with $a_0$ and $b_0$ which can be obtained following the standard thermodynamics of superconductors for $T\lesssim T_c$ which was recently formulated for flat band superconductors~\cite{ktlaw2024ginzburg}. We assume momentum independent and real order parameter. The order parameter is self consistently obatined by solving $\partial \mathcal{F}[\Delta_0]/\partial\Delta_0= 0$ which gives 
\begin{equation}
\label{eq:GL-gap}
    a_0 + b_0 \Delta_0^2 = 0.
\end{equation}
Here, $a_0 = 1/g - \beta^{-1}\sum_{\vec k , n}G_e(\vec k, \omega_n) G_h(\vec k, \omega_n)$ and $b_0 = (\beta^{-1}/2) \sum_{\vec k , n} G_e^2(\vec k, \omega_n) G_h^2(\vec k, \omega_n)$ with $G_{e/h}(\vec k, \omega_n) = (\pm i \omega_n - \epsilon_{\pm \vec k}+\mu)^{-1}$ being the single particle/hole Mastsubara Green's function; $\sum_n$ denoted sum over Matsubara frequency. Performing the Matsubara sums, we can re-write Eq.~(\ref{eq:collect_changegap}) as
\begin{equation}
\label{eq:collect_changegap2}
   \frac{\delta\Delta}{\Delta_0}\left( a_0 + 2b_0 \Delta_0^2\right) =- \sum_{\vec k} \frac{\delta f_{\vec k}}{E_{\vec{k}}} 
\end{equation}
and utilizing Eq.~(\ref{eq:GL-gap}) we obtain 
\begin{equation}
    \frac{\delta\Delta}{\Delta_0} = a_0^{-1} \frac{\delta f_{\vec k}}{E_{\vec{k}}} 
\end{equation}
which is Eq.~(\ref{eq:changegap}) of the main text. Using $1/g = \tanh[\beta_c(\epsilon_{\vec k} - \mu)/2]/2(\epsilon_{\vec k} - \mu)$, for $T\lesssim T_c$ we obtain 
\begin{equation}
    a_0 = \nu_F \left(\frac{T-T_c}{T_c}\right) \tanh\left(\frac{\beta_c W}{4} \right)
\end{equation} 
for an isotropic flat band with $\epsilon_{\vec k} \in [-W/2, W/2]$ satisfying $\beta_c W\lesssim 1$. 

 \subsection{Continuum model for strained TBG-hBN heterostructure}
In this section, we detail how we simulated the electronic structure of TBG using the continuum model. For TBG, we define the lattice structure as in Ref. \cite{arora2022quantum}. In each graphene layer the primitive (original) lattice vectors are $\vec{a}_1 = a_{\rm G}(1,0)$ and $\vec{a}_2 = a_{\rm G}(1/2,\sqrt{3}/2)$ with $a_{\rm G}=0.246$ nm being the lattice constant. The corresponding reciprocal space lattice vectors are $\vec{b}_1 = (2\pi/a_{\rm G})(1,-1/\sqrt{3})$ and $\vec{b}_2 = (2\pi/a_{\rm G})(0,2/\sqrt{3})$, and Dirac points are located at $K_\zeta = -\zeta(2\vec{b}_1 +\vec{b}_2)/3$. For a twist angle $\theta$ (accounting for the rotation of layers), the lattice vectors of layer $l$ are given by $\vec{a}_{l,i} = R(\mp\theta/2)\vec{a}_i$, $\mp$ for $l=1,2$ respectively, and $R(\theta)$ represents rotation by an angle $\theta$ about the normal. Also, from $\vec{a}_{l,i}.\vec{b}_{l',j} = 2\pi \delta_{ij}\delta_{ll'}$ we can check that the reciprocal lattice vectors become $\vec{b}_{l,i} = R(\mp\theta/2)\vec{b}_i$ with corresponding Dirac points now located at $\vec{K}_{l,\zeta} =  -\zeta(2\vec{b}_{l,1} +\vec{b}_{l,2})/3$.

At small angles, the slight mismatch of the lattice period between two layers gives rise to long range moir\'e superlattices. The reciprocal lattice vectors for these moir\'e superlattices are given as $\vec{g}_i = \vec{b}_{1,i} - \vec{b}_{2,i}$. The superlattice vectors $\vec{L}$, can then be found using $\vec{g}_i.\vec{L}_j = 2\pi \delta_{ij}$, where $\vec{L}_1$ and $\vec{L}_2$ span the moir\'e unit cell with lattice constant $L = \vec{L}_1 = \vec{L}_2 = a_{\rm G}/[2\sin\theta/2]$. 

Next, when the moir\'e superlattice constant is much longer than the atomic scale, the electronic structure can be described using an effective continuum model for each valley $\zeta=\pm$. The total Hamiltonian is block diagonal in the valley index, and for each valley, the effective Hamiltonian in the continuum model is written in terms of the sublattice and layer basis $(A_1, B_1, A_2, B_2)$ 
\begin{equation}
H_\zeta = 	\begin{pmatrix}
H_{1,\zeta}(\vec{p}) & T^\dag_\zeta \\
T_\zeta & H_{2,\zeta}(\vec{p})
\end{pmatrix}
\end{equation}
where $H_{l,\zeta} = -\hbar v_F R(\pm \theta/2)\vec{p}.(\zeta\sigma_x, \sigma_y)$ is the Hamiltonian for each layer with $\hbar v_F/a_{\rm G} = 2135.4$ meV, and 
\begin{equation}
T_\zeta = \begin{pmatrix}
t_{\rm AA} & t'_{\rm AB} \\
t'_{\rm AB} & t_{\rm AA} 
\end{pmatrix} +
\begin{pmatrix}
t_{\rm AA} & t'_{\rm AB}e^{-i\zeta\frac{2\pi}{3}} \\
t'_{\rm AB}e^{i\zeta\frac{2\pi}{3}} & t_{\rm AA} 
\end{pmatrix}e^{i\zeta \vec{g}_1.\vec{r}}
+ 
\begin{pmatrix}
t_{\rm AA} & t'_{\rm AB}e^{i\zeta\frac{2\pi}{3}} \\
t'_{\rm AB}e^{-i\zeta\frac{2\pi}{3}} & t_{\rm AA}
\end{pmatrix}e^{i\zeta (\vec{g}_1+\vec{g}_2).\vec{r}}
\end{equation}
where $\vec{g}_i$ is the reciprocal lattice vector of mBZ. We use the tunnelling parameter $t'_{\rm AB}=97.5$ meV, and use different values for $t_{AA}$. When hBN is aligned with the graphene layers, $C_2$ symmetry is broken modifying the layer Hamiltonians $H_{l,\zeta}$. This can be described by introducing a sublattice staggered potential $\Delta_l$ so that the Hamiltonian for each layer $H_{l,\zeta}(\vec{p}) \rightarrow H_{l,\zeta}(\vec{p})+\Delta_l\sigma_z$.

Finally, the presence of a uniaxial heterostrain in TBG of magnitude $\chi$ can be described by the linear strain tensor
\begin{equation}
\label{eq:strain}
\mathcal{E}_l = \mathcal{F}(l)\chi\begin{pmatrix}
-\cos^2\varphi + \nu \sin^2\varphi & (1+ \nu) \cos\varphi\sin\varphi \\
(1+ \nu) \cos\varphi\sin\varphi & \nu\cos^2\varphi - \sin^2\varphi
\end{pmatrix}
\end{equation}
where $\mathcal{F}(l=1,2)=\mp1/2$, $\nu=0.165$ is the Poisson ratio of graphene and $\varphi$ gives direction of the applied strain. The strain tensor satisfies general transformations in each layer, $\vec{a}_l \rightarrow \vec{a}_l'= [1+\mathcal{E}_l]\vec{a}_l$ and $\vec{b}_l \rightarrow \vec{b}_l ' \approx [1-\mathcal{E}_l^T]\vec{b}_l$ for real and reciprocal lattice vectors respectively. The strain induced geometric deformations affect the interlayer coupling and further changes the electron motion via gauge field $\vec{A}_l = \sqrt{3}\beta/2a_{\rm G}(\mathcal{E}^{xx}_l+\mathcal{E}^{yy}_l, -2\mathcal{E}^{xy}_l )$, where $\beta = 3.14$. As a result, we have $\vec{p}\rightarrow \vec{p}_{l,\zeta} = [1+\mathcal{E}_l^T][\vec{k}-\mathcal{K}_{l,\zeta}]$ with $\mathcal{K}_{l,\zeta} = [1-\mathcal{E}_l^T]\vec{K}_{l,\zeta} - \zeta \vec{A}_l$. 

The effective TBG Hamiltonian modified by the effects of strain and hBN alignment with graphene layers via sublattice staggered potential, can be re-written as 
\begin{equation}
\label{eq:tbghamiltonianstrainandhbn}
\mathcal{H}_\zeta = \begin{pmatrix}
H_{1,\zeta}(\vec{p}_{1,\zeta})+\Delta_1\sigma_z & T_\zeta^\dag \\
T_\zeta & H_{2,\zeta}(\vec{p}_{2,\zeta})+\Delta_2\sigma_z
\end{pmatrix}
\end{equation}
Note that for a given $\vec{q}$ in the mBZ, the 4$\times$4 Hamiltonian in Eq. (\ref{eq:tbghamiltonianstrainandhbn}) is cast into a multiband eigensystem problem as the interlayer coupling leads to hybridisation of the eigenstates at Bloch vectors $\vec{q}$ and $\vec{p}'=\vec{p}+\vec{g}$, where $\vec{g} = m_1\vec{g}_1 +m_2\vec{g}_2$ and $m_{1,2}\in\mathbb{Z}$. We truncate the size of the matrix by defining a circular cut-off $|\vec{p}-\vec{p}'|<4|\vec{g}_1|$. For a given Bloch vector $\vec{p}$, this gives us 61 sites in reciprocal space, and a corresponding matrix of size $244 \times 244$ which is then diagonalized to obtain eigenvalues and eigenvectors. TBG bandstructure with strain and hBN encapsulation is shown in Fig. at different values of $t_{AA}$.

\begin{figure}
    \centering
    \includegraphics[width=0.3\linewidth]{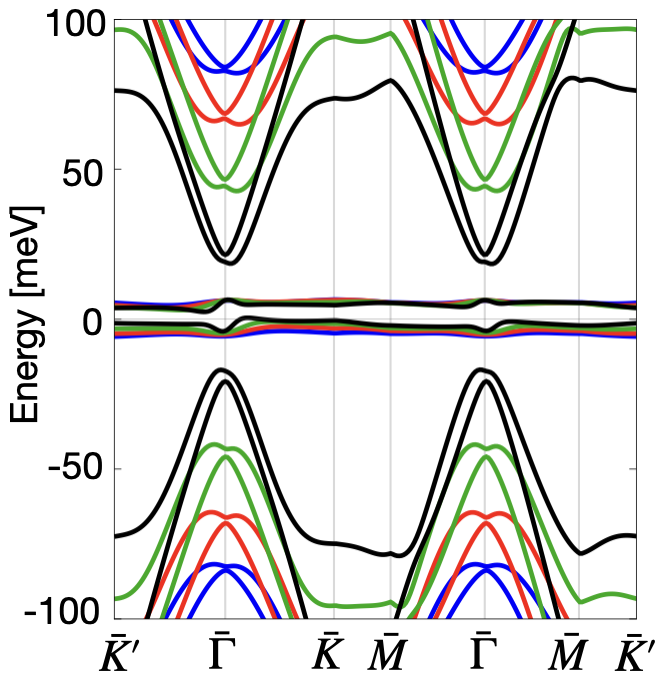}
    \caption{TBG bandstructure for $t_{AA}=20$ meV (blue), $t_{AA}=40$ meV (red), $t_{AA}=60$ meV (green) and $t_{AA}=80$ meV (black). Additionally, we use $\delta_1=\delta_2=5$ meV and strain of 0.1\%.}
    \label{fig:enter-label}
\end{figure}

\subsection{Numerical evaluation of $\delta\Delta/\Delta_0$ in TBG-heterostructure}
The change in superconducting order $\delta\Delta/\Delta_0$ is calculated using Eq.~(\ref{eq:changegap}) where the change in distribution function under drive is calculated using collision integral in Eq.~(\ref{eq:collision_integral}). Two Riemann sums, over $\vec k'$ and $\vec k$ in Eq.~(\ref{eq:collision_integral}) and Eq.~(\ref{eq:changegap}), respectively are calculated over discrete grid in $(k_x, k_y)$ plane of mBZ. Two sums make the problem computationally heavy for which we choose a coarse grid of $50 \times 50$ points, and include only two flat bands with the nearest remote bands in our calculation. Further, we choose $\Delta_0 = 0.1$ meV and approximate $\delta$-function as a Lorentzian with phenomenological energy broadening of 0.1 meV.  
\end{document}